\documentclass[aps]{revtex4}
\begin{document}
\title{LANDAU SINGULARITY AND THE INSTABILITY OF VACUUM STATE IN QED}
\author{Mofazzal Azam}
\address{Theoretical Physics Division\\
         Bhabha Atomic Research Centre\\
         Mumbai-400085, India}

\begin{abstract}
Quantum Eletrodynamics (QED) is considered as the most successful
of all physical theories. It can predict numerical values of physical
quantities to a spectacular degree of accuracy. However, from the
very early days it has been known that, in QED, there are two
important problems which are linked with the very foundation of
the theory. In 1952, Dyson put forward strong argumnts to suggest that
the perturbation seires in quantum electrodynamics can not be convergent.
Just three years latter, in 1955, Landau argued that the effective running
coupling constant in QED has a pole (Landau singularity) albeit at some
very high energy scale. This paper addresses, in details, the question of
stability of perturbative vacuum state of QED in the light of these two
well known problems. \\
Landau has been a cult-like figure for many of us who studied theoretical
physics in the former Soviet Union.
As an undergraduate student in the department of theoretical physics of
People's Friendship University, Moscow, in 1970's, I grew up hearing
colourful stories about the legendary persona of Lev Davidovich
from my teachers some of whom had known him at personal level.
It is a great honour for me to contribute this article to the
Landau centenary volume of Eletronics Journal of Theoretical Physics.
\end{abstract}
\maketitle
\vskip .8 in
\par Landau singularity of the effective coupling constant in
quantum electrodynamics has been known for more than half a
century \cite{landau}. The original derivation of the singularity
was based on the summation of one loop diagrams of vacuum
polarization tensor for photons in the perturbation theory.
In the early days the validity of such an expansion was looked upon
sceptically by many people including the authours  themselves.
However, after the advent of $~1/N~$ expansion technique by t'~Hooft,
it was soon realised that this singularity appears at the
leading order in $1/N_{f}$ expansion where $N_{f}$ is the
number of specieses of electrons, also called the number of flavours.
This implies that in the infinite flavour limit this singularity is
exact provided the perturbation series in $~1/N_{f}~$ expansion converges.
This issue of convergence of the perturbation series will be one
of the central themes latter in the paper. There have
been attempts to interpret this singularity in many ways.
Landau and Pomeranchuk tried  to argue that this singularity
reflects the fact that at short distances strong vacuum polarization
effects screen the electric charge completely \cite{landau}.
Others, including Shirkov, have called it
Landau ghost reflecting the internal inconsistency of
quantum electrodynamics \cite{shirkov}.
\par It should be mentioned here that the stability of ground state
and the possible existence of an ultra-violet fixed point has been studied
extensively in the lattice formulation of QED in the wide range of value
of the fermion flavour $N_f$ by Kogut {\it et al} \cite{kogut1,kogut2}.
Using lattice formulation, Kim {\it et al} has argued for the triviality
of QED \cite{kogut}. On the other hand in
massless QED in the continuum, Miransky  argues that there exists a
chiral symmetry breaking phase \cite{mir}.
However, in these studies Landau
singularity plays no role. There has been some recent studies of Landau
singularity using lattice formulation of QED by Gockeler {\it et al}
(\cite{go} and references there in).
These studies seem to suggest that chiral symmetry breaking allows
QED to escape Landau singularity. But then chiral symmetry breaking,
as their study shows, seems to be intimately connected
with trivility of QED. Landau singularity has also been considered
from a different perspective by Gies and Jaeckel \cite{gies}, and
by Langfeld {\it et al} \cite{lang}. The details of these approaches
can be found in the cited references and will not be discussed here.
Our approach will be very different from the ones described in the
publications above. We will rather be interested
in finding the meaning of Landau singularity than finding a way
to escape it.\\
We want to consider the issue of stability  of "ground state/vacuum state"
of quantum theory in the unified and broad perspective of
quantum field theory and the many-body theory. Therefore,
at first, we consider many-body ground state of two purely quantum
mechanical systems: "Coulomb system with large number of
flavours of fermions" and "System of weakly interacting electron gas
in a condensed matter system". These are discussed in section-I
and section-II. After this, we return
to the main theme of the paper. In section-III, we reproduce Dyson's
argument for the divergence of perturbation series in QED. In
section-IV, we show the connection between the Landau singularity and
instability of vacuum state in QED. It is argued in the Dysonian
framework, how the divergence of the series removes both these problem.
In section-V, we explain how a divergent asymptotic series can
give rise to physically meaningful physical quantities. In the last and
concluding section, we comment on the non-perturbative
aspects of QED. Dyson's original arguments, as well as, our
studies of Landau singularity and vacuum state shows that
the physical observables in non-perturbative QED
are non-analytic in the coupling , $e^2$, as well as the inverse
flavour, $1/N_f$, and the perturbative power series in these parameters
can not capture this behaviour. In the absence of non-perturbative
theory, we suggest that there should be attempts to experimentally
search for non-analyticity in some physical observables in QED.

\section{Ground state of Coulomb system with large number of flavours
of fermions}

\par We will be investigating in this paper the question of stability
of vacuum state in quantum electrodynamics. In this context,
it is interesting to have some information regarding
the stability of quantum machanical ground state
of a large system of charged particle.
Many-body theory of Coulomb system of fermions have been studied
extensively in several publications.
In quantum theory  a many-body system with ground
state energy $~E_0(N)~$ is called thermodynamically stable
or simply stable if $ ~E_0(N)/N ~$ is bounded below
when the number of particles, $~N\rightarrow \infty~$.
In this context, the question of
stability of matter consisting of negatively charged electrons
and positively charged nuclei is very important. It was in 1967 when
Dyson and Lenard proved  that, in the framework of nonrelativitic
quantum mechanics, matter consisting of N electrons and K static
point nuclei of charge $~Z~$ ($=N/K$) is stable \cite{dyson}.
Subsequently, Lieb and his collaborators have made a very detailed
investigation of the stability of matter in nonrelativistic as well as
relativistic case ( \cite{lieb3,lieb1} and references there in ).
These studies seem to suggest that at
high enough energies quantum eletrodynamics may not be a well
behaved theory.
Thermodynamic stability of a system of particles interacting via
Coulomb interaction is associated with the control of the
short distance behaviour of the interaction. In the nonrelativistic
limit, zero point kinetic energy of the fermions controls this short
distance behavoiur. However, in the relativistic limit, there is
a need for certain bounds on the value of the fine structure constant
\cite{lieb3,lieb1,conlon}. Landau singularity is also believed to be
associated with the "high energy / short distance"
behaviour of the electromagnetic interaction \cite{landau}.
\par Let us consider the large flavour case in some detail.
In 3-dimensional space, let us consider a volume
of linear dimension $~R~$ where there are $~N~$
number of positively charged and $~N~$ number of negatively charged
fermions. Fermions of both positive and negative charges come
with $~N_f~$ flavours. Particles are assumed to be of the same mass.
We assume that $~N~>>~N_f~$, however, both the number of particles
and number of flavours are assumed to be large. Taking into
account the the Pauli exclusion principle for the case
involoving multiflavour fermions in $3$-dimensional space,
the kinetic energy K, apart from some numerical factors,
can be written \cite{yau}
as $~~K \approx \frac{\hbar^2}{2m} \frac{N^{5/3}}{R^{2}N_{f}^{2/3}}$.
The potential energy $U$ due to Coulomb interaction
is given by \cite{lieb3},
$~~U=-a_0 \frac{e^{2}N^{4/3}}{R}$,$~~a_0>0~~$ is a  numerical constant.
The total energy, $E(R)$, is given by
\begin{eqnarray}
E(R)=K+U \approx \frac{\hbar^2}{2m} \frac{N^{5/3}}{R^{2}N_{f}^{2/3}}
-a_0 \frac{e^{2}N^{4/3}}{R}
\end{eqnarray}
The ground state is obtained by minimizing the total energy,
$E(R)$, with respect to $R$.
\begin{eqnarray}
E_0~~\approx~ -\frac{ma_{0}^{2}}{\hbar^{2}}(e^2 N_f^{1/3})^2 N ~;~~~
\epsilon_0 ~=~E_0/N \approx~ -\frac{ma_{0}^{2}}
{\hbar^{2}}(e^2 N_f^{1/3})^2
\nonumber\\
R_0~~\approx~\frac{\hbar^2}{ma_0}
(\frac{N}{N_f})^{1/3}\frac{1}{e^2 N_f^{1/3}}~;~~~
r_0=\frac{R_0}{(N/N_f)^{1/3}} \approx
~\frac{\hbar^2}{ma_0}\frac{1}{e^2 N_f^{1/3}}
\end{eqnarray}
When the number of flavour $~N_f~$ increases, the size of the small box
$~r_0~$ decreases and when it approaches Compton wave length
$~\hbar/mc~$, we can no longer use the nonrelativistic quantum
mechanics. Therefore, the above estimates are correct only when
\begin{eqnarray}
N_f~<<~ \frac{1}{a_{0}^3} \frac{1}{(\frac{e^2}{\hbar c})^3}
\end{eqnarray}
In the relativistic quantum theory of many-body systems, it is
almost customary now to take $~c|p|~$ \cite{lieb3},
where $c$ is the velocity of light,
as the kinetic energy of
individual particles. Therefore, the kinetic energy $K$ for the system
of $N$ particles with $~N_f~$ flavour in a region with linear dimension
$R$ is given by, $K~=~(\hbar c N^{4/3})/(R N_{f}^{1/3})~~$.
Thus, in the relativistic case,
\begin{eqnarray}
E(R)=~K~+~U~\approx~\frac{\hbar c N^{4/3}}{R N_{f}^{1/3}}~
-a_0 \frac{e^{2}N^{4/3}}{R}
\end{eqnarray}
In this case stability of ground state requires $E(R)~\geq~0$.This
leads to the condition,
\begin{eqnarray}
N_f~\leq~\frac{1}{a_{0}^3} \frac{1}{(\frac{e^2}{\hbar c})^3}
\end{eqnarray}
which in the unit $~c~=\hbar~=1~$, becomes
$N_f~ \leq~~\frac{1}{(a_{0}e^{2})^3}~$.
At this ponit a few comments are necessary. We considered a theory
with large $~N~$ and large $~N_f~$, and it is implicit in our discussion
that we are considering, $~N~\rightarrow\infty~$, but we have not said
much about the flavour $~N_f~$. From the equations above,
we find that it is also possible to take the limit,
$~N_f~\rightarrow~ \infty~$ but slower than N. Assuming that
in the limit $~N_f~\rightarrow~\infty~$ and $~e~\rightarrow~0~~$,
$e^{2} N_{f}^{1/3}~=constant$, and that it satisfies equation Eq.(5),
it is easy to conclude that, in this limit, matter is stable.
This mathematical limit, has the
physical implication that when the charge, $e$ is small and
the flavour $N_f$ is large, the
relevant parameter that sets the stability criteria
is $~e^2 N_{f}^{1/3}~$. It is the value of
$~e^2 N_{f}^{1/3}~$   that decides the stability of the system. When
$~e^2 N_{f}^{1/3}~>~\frac{1}{a_{0}^{3}}~$, in other words, when
the number of flavours
$N_f~>~ \frac{1}{(a_{0}e^{2})^3}~$,
the many-body system is thermodynamically unstable.\\
In subsequent section, we will find that in quantum
electrodynamics the relevant parameter is not
$e^2 N_{f}^{1/3}$ but $e^2 N_{f}$.

\section{Feldman Model of Weakly Interacting Electron Gas}

The main theme of this paper, as annouced in the abstract, is
to look for the meaning of Landau singularity of the effective
coupling constant in QED. There are some other contexts
in which the effective coupling constant develops singularity.
We have in mind some condensed matter systems, where these
singularities have well defined physical meaning.
In this section, we refer to the renormalisation group analysis of weakly
interacting Fermi system with short range potential at
finite density and zero temparature by Feldman {\it et al}
\cite{feld,froh,shan}. The iterative
renormalization group transformations
show that if there is an attractive interaction among the electrons
in any angular momentum channel, then there appears similar type of
singularity in some suitably defined running coupling constant.
We briefly describe here the Feldman model of weakly interacting
electron gas. This section is essentially a short description
of the results taken from the very detailed review paper by
Froehlich, Chen and Seifert  \cite{froh}.\\
The model of weakly interacting electron gas studied by
Feldman {\it et al}
is a condensed matter Fermi system in thermal equillibrium
at some temparature $T$ (for simplicity, assume $T=0$)
and chemical potential $\mu$.
On microscopic
scale($\approx 10^{-8}$ cm), it can be described approximately
in terms of non-relativistic electrons with
short range two body interactions.
The thermodynamic quantities such as conductivity depend only on
physical properties of the system at mesoscopic length scale
($\approx 10^{-4}$ cm),
and therefore, are determined from processes involving momenta
of the order of
$\frac{k_{F}}{\lambda}$ around the Fermi surface, where
the parameter, $\lambda >>1$,  should be thought of
as a ratio of meso-to-microscopic length scale. This is
generally refered to as
the scaling limit(large $\lambda$, low frequencies) of the system.
The most important observation of Feldman et. al.
is that in the scaling
limit, systems of non-relativistic (free)
electrons in $d$ spatial dimensions behave like a system of
multi-flavoured relativistic chiral Dirac fermions
in $1+1$ dimensions.The number of flavours
$N \approx ~const. ~ \lambda^{d-1}$.
It is possible then to set up a renormalization group
improved perturbation
theory in $\frac{1}{\lambda}$ around the non-interacting electron gas,
where in, the large number of flavours $N$, play an important role
in actual calculations.
\subsection{Free Electron gas and the Multiflavour Relativistic
Fermions in $1+1$ Dimensions}
Let us consider a system of non-relativistic
free electrons in $d$ spatial dimensions with the Euclidean action,

\begin{eqnarray}
S_{0}(\psi^*,\psi)=\sum_{\sigma} \int d^{d+1} x \psi_{\sigma}^*(x)
(\rm{i} \partial_0-\frac{1}{2m} \Delta -\mu)\psi_{\sigma}(x)
\end{eqnarray}
The Euclidean free fermion Green's function, $ G^{0}_{\sigma \sigma'}
(x-y)$,
where $\sigma$ and $\sigma'$  are the spin indices,
$x=(t,\vec{x})$ and $y=(s,\vec{y})$,
$t$ and $s$ are imaginary times,
$t>s$, is given by,

\begin{eqnarray}
G^{0}_{\sigma \sigma'}(x-y)=\langle \psi_{\sigma}^*(x)\psi_{\sigma}(y)
\rangle_{\mu}
=-\delta_{\sigma \sigma'} \int (dk)\frac{e^{-ik_0(t-s)+i\vec{k}
(\vec{x}- \vec{y})}}{ik_0 -(\frac{k^2}{2m}-\mu)}
\end{eqnarray}
Where we have used $(dk)=\frac{1}{(2\pi)^(d+1)} d^{d+1}k$.
In the scaling limit, the leading contributions to
$ G^{0}_{\sigma \sigma'}(x-y)$
come from modes whose momenta are contained in a
shell $S_{F}^{(\lambda)}$ of thickness $\frac{k_{F}}{\lambda}$
around the Fermi surface $S_{F}$.
In order to approximate the Green's function, let us
introduce the new variables
$\vec{\omega}, ~p_{\parallel}, ~p_{0}$ such that $k_{F}\vec{\omega}
\in S_F$, $p_0 =k_0$ and $\vec{k}=(k_F+p_{\parallel})\vec{\omega}$.
If $\vec{k}\in S_{F}^{(\lambda)}$, then $p_{\parallel}<<k_{F}$, and
we can approximate the integrand of Eq.(7), by dropping
$p_{\parallel}^{2}$ term in the denominator.In other words,
\begin{eqnarray}
G^{0}_{\sigma \sigma'}(x-y)= \delta_{\sigma \sigma'}\int
\frac{d {\bf \sigma} (\vec{\omega})}{(2\pi)^{d-1}} k_{F}^{d-1} e^{ik_{F}
\vec{\omega}(\vec{x}-\vec{y})} G_{c}(t-s, \vec{\omega}(\vec{x}-\vec{y})
\end{eqnarray}
where $d {\bf \sigma}(\vec{\omega})$ is the uniform measure on unit
sphere and
\begin{eqnarray}
G_{c}(t-s, \vec{\omega}(\vec{x}-\vec{y}))=-\int \frac{dp_{0}}{2\pi}
\frac{dp_{\parallel}}{2\pi} \frac{e^{-i k_0 (t-s)+ip_{\parallel}
\vec{\omega}(\vec{x}- \vec{y})}}{ip_0 -v_{F}p_{\parallel}}
\end{eqnarray}
is the Green's function of chiral Dirac fermion in $1+1$ dimension.
$v_{F}=k_{F}/m$ is the  Fermi velocity.The $\vec{\omega}$-integration
in Eq.(8) can be further approximated by replacing it with summation
over discrete directions $\vec{\omega}_{j}$ by dividing the shell
$S_{F}^{(\lambda)}$ into $N$ small boxes $B_{\vec{\omega}_{j}}, j=1,..,N$
of roughly cubical shape.The box, $B_{\vec{\omega}_{j}}$, is centered at
$\vec{\omega}_{j} \in S_F$ and has an approximate side length
$\frac{k_F}{\lambda}$.The number of boxes, $N=\Omega_{d-1} \lambda ^{d-1}$,
where $\Omega_{d-1}$ is the surface volume of unit sphere in
$d$ spatial dimensions.The Green's function is, now, given by
\begin{eqnarray}
G^{0}_{\sigma \sigma'}(x-y)=-\delta_{\sigma \sigma'}\sum_{\vec{\omega}_{j}}
\int \frac{dp_{0}}{2\pi} \frac{dp_{\parallel}}{2\pi} \frac{p_{\perp}}
{2\pi}
\frac{e^{-ip_0(t-s)+i\vec{p} (\vec{x}- \vec{y})}}
{ip_0 -v_{F}p_{\parallel}}
\end{eqnarray}
where $\vec{p}=p_{\parallel}\vec{\omega}+\vec{p}_{\perp}$ is a vector
in $B_{\vec{\omega}_{j}}-k_{F} \vec{\omega}_{j}$ and $p_{0}\in
\mathcal{R}$.
Thus in the scaling limit, the behaviour of a $d$-dimensional
non-relastivistic free electron gas is described by
$N=\Omega_{d-1} \lambda ^{d-1}$ flavours of free chiral Dirac fermions
in $1+1$ dimensional space-time.The propagator
$G_{c}(t-s, \vec{\omega}(\vec{x}-\vec{y})$ depends on the flavour
index $ ~\vec{\omega} ~$.But the energy of an electron or hole with momenta
$ ~\vec{k} ~$ , depends only on $p_{\parallel}$, where
$ ~p_{\parallel}=\vec{k}.\vec{\omega}-k_F ~$ and
$ ~\vec{\omega}=\frac{\vec{k}}{|k|} ~$.It is proportional to
$ ~p_{\parallel} ~$ , just as for relativistic fermions in
$1+1$ dimensions.
\subsection{Renormalization Group Flow and the BCS Instability}
Large scale behaviour of the weakly interacting system is
described by an effective action.
To see how the effective action is calculated,
let us consider a system with Euclidean action of the form,
\begin{eqnarray}
S(\psi^*,\psi)=S_{0}(\psi^*,\psi) + S_{I}(\psi^*,\psi)
\end{eqnarray}
where $S_{0}(\psi^*,\psi)$ is the quadratic part given in Eq.(3)
and $S_{I}(\psi^*,\psi)$ is the quartic interaction term.For a weakly
interacting electron gas,
\begin{eqnarray}
S_{I}(\psi^*,\psi)=g_{0}k_{F}^{1-d}\sum_{\sigma,\sigma'}\int d^{d+1}x
d^{d+1}y :\psi_{\sigma}^{*}(x) \psi_{\sigma}(x) v(\vec{x}-\vec{y})
\delta(x_0-y_0) \psi_{\sigma'}^{*}(y) \psi_{\sigma'}(y)
\end{eqnarray}
The factor $k_{F}^{1-d}$ ensures that $g_{0}$ is
dimensionless. The two body potential $v(\vec{x})$ is
assumed to be smooth and short range, and therefore, its Fourier
transform $\hat{v}(\vec{k})$ is also smooth.
In order to calculate the effective actions,
we first split the field variable into  slow and fast modes,
$ ~\hat{\psi}=\hat{\psi}_{<}+\hat{\psi}_{>} ~$, where
Supp $\hat{\psi}_{>} \subset R \times
(R^d ~ \setminus ~
S_{F}^{(\lambda)}) ~~(\rm{region} >) ~~$ and
Supp $\hat{\psi}_{<} \subset R \times S_{F}^{(\lambda)}
 ~~(\rm{region} <) ~~$.
We then integrate out the fast modes using the functional integral
for fermions,
\begin{eqnarray}
e^{-S_{eff}(\hat{\psi}^{*}_{<},\hat{\psi}_{<})}=
\frac{1}{\Xi}\int D\hat{\psi}^{*}_{>} D\hat{\psi}_{>}
e^{-S(\hat{\psi}^{*}_{>},\hat{\psi}^{*}_{<},\hat{\psi}_{>},\hat{\psi}_{<})}
\end{eqnarray}
Now using the linked cluster theorem, we obtain
\begin{eqnarray}
e^{-S_{eff}(\hat{\psi}^{*}_{<},\hat{\psi}_{<})}=
\exp{(-S_{0,<}-<S_{I}>_{G^{0}_{>}}+\frac{1}{2}<S_{I};S_{I}>_{G^{0}_{>}}
-\frac{1}{3}<S_{I};S_{I};S_{I}>_{G^{0}_{>}}-...)}
\end{eqnarray}
The abbreviations, $<a;b>$ $<a;b;c>$ etc., denote
the connected correlators.
The subscripts $G^{0}_{>}$ indicate that the
expectations $<(.)>_{G^{0}_{>}}$ are calculated using infrared cut-off
free propagators  in accordance with the functional measure,
\begin{eqnarray}
dP(\hat{\psi}^{*}_{>},\hat{\psi}_{>})
=(1/\Xi) D\hat{\psi}^{*}_{>} D\hat{\psi}_{>}
e^{-S_{0}(\hat{\psi}^{*}_{>},\hat{\psi}_{>})} \nonumber
\end{eqnarray}
The connected correlators can be evaluated using the Feynman
diagram technique. Therefore, the effective action can be calculated
once we know the amplitudes of connected Feynman diagrams.
It is clear that
the $S_{eff}$ contains far more interactions than the original
quartic interaction $S_{I}$.However, for weakly interacting systems
the original coupling remains dominant.
To carry out the iterative
renormalization group  scheme of Feldman {\it et al}, we
choose some large scale $\lambda_{0}<< \frac{1}{\sqrt{g_0}}$ and
calculate the effective action, $S_{eff}$ perturbatively to leading order
in $\frac{1}{\lambda_0}$.The effective action depends on modes
corresponding to wave vectors located in the shell, $S_{F}^{(\lambda)}$,
of width
$\frac{k_F}{\lambda _0}$ around the Fermi surface.
Now, we divide the shell
$S_{F}^{(\lambda)}$ into $N=const. ~ \lambda^{d-1}$ cubical
boxes, $B_{\vec{\omega}_{j}}$,
of approximate  side length
$\frac{k_F}{\lambda _0}$.Next, we rescale all the momenta so that,
instead of belonging to the boxes $B_{\vec{\omega}_{j}}$
they are contained in boxes
$\tilde{B}_{\vec{\omega}_{j}}$ of side length $\approx k_{F}$.
These two steps are generally known as decimation of degrees of
freedom and rescaling.
The renormalization group scheme
consists of iteration of these two steps.
\par Assume that the degrees of freedom corresponding to momenta
not lying in $S_{F}^{(\lambda)}$ have been integrated out.
Let $\hat{\psi}_{\sigma}(k), k\in R \times S_{F}^{(\lambda)}$,
denote these modes. The sector fields are defined as
\begin{eqnarray}
\psi_{\vec{\omega},\sigma}(x)=\int_{R \times B_{\vec{\omega}}}(dk)
e^{i(k_{0}t-(\vec{k}-k_{F}\vec{\omega})\vec{x})} \hat{\psi}_{\sigma}(k)
\end{eqnarray}
It is easy to see that
$ ~\psi_{\sigma}(x)=\sum_{\vec{\omega}} e^{\rm{i} k_{F}\vec{\omega}\vec{x}}
\hat{\psi}_{\vec{\omega},\sigma}(x) ~$.
Inserting the Fourier transform of sector fields in Eq.(6) and Eq.(12),
and carrying out some algebraic manipulations, we obtain
\begin{eqnarray}
S_{0}=-\sum_{\vec{\omega},\sigma}\int_
{\mathcal{R}\times (B_{\vec{\omega}}-k_{F}\vec{\omega})}(dp)
\hat{\psi}^{*} _{\vec{\omega},\sigma}(p)\big(\rm{i}p_{0}-v_{F}\vec{\omega}
\vec{p} +O\big(\frac{1}{\lambda^{2}}\big)\big)
\hat{\psi}_{\vec{\omega},\sigma}(p)
\end{eqnarray}
\begin{eqnarray}
S_{I}&=&\frac{g_{0}}{2} k_{F}^{1-d} \sum_{\vec{\omega}_1,..,\vec{\omega}_4
;\sigma,\sigma'}\hat{v}(k_F(\vec{\omega}_1- \vec{\omega}_4)) \int
(dp_{1})...(dp_{4})(2\pi)^{d+1} \delta (p_1+p_2- p_3-p_4)  \nonumber\\
& &\hat{\psi}^{*} _{\vec{\omega}_1, \sigma}(p_1)
\hat{\psi}^{*} _{\vec{\omega}_2,\sigma'}(p_2)
\hat{\psi}_{\vec{\omega}_3,\sigma'}(p_3)
\hat{\psi}_{\vec{\omega}_4,\sigma}(p_4) +
\rm{terms ~~of ~~higher ~~order
 ~~in}  \frac{1}{\lambda}
\end{eqnarray}
Using cluster expansions to integrate out the degrees of freedom
corresponding to
momenta outside the shell $S_{F}^{(\lambda_0)}$, one can show
that , at scale $\frac{k_F}{\lambda}$ , the effective action has the
form given by Eq.(13) and Eq.(14) except that
$\hat{v}(k_F(\vec{\omega}_1- \vec{\omega}_4))$ is replaced by
a coupling function
$g(\vec{\omega}_1,\vec{\omega}_2,\vec{\omega}_3,\vec{\omega}_4)
\approx \hat{v}(k_F(\vec{\omega}_1- \vec{\omega}_4))$ with
$\vec{\omega}_1 + \vec{\omega}_2 = \vec{\omega}_3 + \vec{\omega}_4$.
\par Next, one considers the rescaling of the fields and the action.The
fields are rescaled in such a that the supports of the Fourier
transformed "sector fields" are boxes,
$\tilde{B}_{\vec{\omega}}=\lambda (B_{\vec{\omega}}-k_{F}\vec{\omega})$,
of roughly cubical shape with
sides of length $k_{F}$, and such that the quadratic part of the action
remains unchanged to leading order in $\frac{1}{\lambda}$. The first
condition implies that
$p \longmapsto \tilde{p}=p\lambda$ and $x \longmapsto \xi=
\frac {x}{\lambda}$.
The rescaled
sector fields and their Fourier transforms are given by,
\begin{eqnarray}
\tilde{\psi}_{\vec{\omega},\sigma}(\xi)=\lambda^{\alpha}
\psi_{\vec{\omega},\sigma}(\lambda \xi) ~~~;~~~
\hat{\tilde{\psi}}_{\vec{\omega},\sigma}(\tilde{p})=\lambda^{\alpha-d-1}
\hat{\psi}_{\vec{\omega},\sigma}(\frac{\tilde{p}}{\lambda})
\end{eqnarray}
Inserting the scaled Fourier transformed fields into quadratic part of
action $S_{0}$, it is easy to see that $S_{0}$ remains unchanged if
the scaling dimension $\alpha=\frac{d}{2}$.
Now, inserting the rescaled fields in the quartic part of the action,
we find that the quartic part has scaling dimension $(1-d)$.
The quartic terms of higher degree in momenta as well as terms of
higher degree in fields appearing in the effective action have
smaller scaling dimensions.
Thus the effective action in terms of scaled sector fields
is,
\begin{eqnarray}
S_{eff}&=&\sum_{\vec{\omega},\sigma}\int (d \tilde{p})
\hat{\tilde{\psi}}^{*}_{\vec{\omega},\sigma}(\tilde{p})
(\rm{i}\tilde{p}_{0}-v_{F}\vec{\omega}
\vec{\tilde{p}})
\hat{\tilde{\psi}}_{\vec{\omega},\sigma}(\tilde{p})
+\frac{1}{2} \frac{1}{\lambda^{d-1}}
\sum_{\vec{\omega}_1+\vec{\omega}_2=\vec{\omega}_3+\vec{\omega}_4
; \sigma,\sigma'}
g(\vec{\omega}_1,\vec{\omega}_2,\vec{\omega}_3,\vec{\omega}_4) \nonumber\\
& &\int
(d\tilde{p}_{1}). \ldots (d\tilde{p}_{4})(2\pi)^{d+1}
\delta (\tilde{p}_1+\tilde{p}_2- \tilde{p}_3-\tilde{p}_4)
\hat{\tilde{\psi}}^{*} _{\vec{\omega}_1, \sigma}(\tilde{p}_1)
\hat{\tilde{\psi}}^{*} _{\vec{\omega}_2,\sigma'}(\tilde{p}_2)
\hat{\tilde{\psi}}_{\vec{\omega}_3,\sigma'}(\tilde{p}_3)
\hat{\tilde{\psi}}_{\vec{\omega}_4,\sigma}(\tilde{p}_4) \nonumber\\
& &+\rm{terms ~~of ~~higher ~~order ~~in}  \frac{1}{\lambda}
\end{eqnarray}
We see that that the inverse propagator for the sector field is
diagonal in $\vec{\omega}$ ,and it depends only on $p_0$ and
$p_{\parallel} =\vec{\omega}\vec{p}$ but not on
$p_{\perp} =\vec{p}-(\vec{\omega}.\vec{p})\vec{\omega}$.
\par We are interested in the renormalization group
flow equations in the leading order in $\frac{1}{\lambda}$.
Therefore, for carrying out the decimation of degrees of freedom,
we will be interested only in those diagrams that contribute
to the amplitude in the leading order in $\frac{1}{\lambda}$.Before
we consider these diagrams, let us consider the possible inter sector
scattering geometries.How many independent
$g^{(0)}(\vec{\omega}_1,\vec{\omega}_2,\vec{\omega}_3,\vec{\omega}_4)$
exists?
For $d=3$, suppose $\vec{\omega}_3 \neq -\vec{\omega}_4$.On the unit
sphere, there  are $N^{(0)}=Const.\lambda_{0}^{d-1}$
different $\vec{\omega}$'s.
But all choises of $\vec{\omega}_1$ and $\vec{\omega}_2$ with
$\vec{\omega}_1 + \vec{\omega}_2 = \vec{\omega}_3 + \vec{\omega}_4$ lie
on a cone containing $\vec{\omega}_3$  and $\vec{\omega}_4$
with $\vec{\omega}_3 + \vec{\omega}_4$ as the symmetry axis.Therefore,
there are $O(\lambda_{0}^{d-2})$ choices.Only when
$\vec{\omega}_3 = - \vec{\omega}_4$ that there are
$N^{(0)}=Const.\lambda_{0}^{d-1}$ choices.
Couplings involving incoming states with
$\vec{\omega}_3 \neq -\vec{\omega}_4$ will be represented by
$g^{(0)}(\vec{\omega}_1,\vec{\omega}_2,\vec{\omega}_3,\vec{\omega}_4)$.
Couplings that involve sectors $\vec{\omega}_3 = -\vec{\omega}_4$
or equivalently $\vec{\omega}_1 = -\vec{\omega}_2$ will be
denoted by $g^{(0)}_{BCS}(\vec{\omega}_1,\vec{\omega}_4)$. Because of
rotational invariance, $g^{(0)}_{BCS}(\vec{\omega}_1,\vec{\omega}_4)$
is a function of only the angle between $\vec{\omega}_1$
and $\vec{\omega}_4$.
\par The chemical potential receives corrections
from connected diagrams with two external electron lines (self energy
correction of the electrons).
The contribution of order zero in
$\frac{1}{\lambda}$ comes only from the
{\bf tadpole} and {\bf turtle} diagrams.
These diagrams contain one internal interaction squiggle of order
$\frac{1}{\lambda_{0}^{d-1}}$ and there are
$N^{(0)}=Const.\lambda_{0}^{d-1}$
choices of the inner particle sector.Therefore, the contribution
is of order zero in $\frac{1}{\lambda}$. The amplitude corresponding
to these tadpole and turtle diagrams turns out to be $p$-independent but
of order $O(g^(0))$.Thus, $ ~\delta \mu_{1}= ~O(g^{(0)}/\lambda_{0}) ~$,
and the renormalized electron propagator is
$ ~G^{ ~R ~}_{\vec{\omega}}(p_{0}, \vec{p}.\vec{\omega}) ~
=- ~[\rm{i}p_{0}-v_{F}\vec{p}.\vec{\omega}+\lambda_{0}
\delta \mu_{1}]^{-1}$.
\par To find the evolution of the coupling constant
$g(\vec{\omega}_1,\vec{\omega}_2,\vec{\omega}_3,\vec{\omega}_4)$,
we have to calculate amplitude of diagrams with four external legs.
It is found that when
$\vec{\omega}_1 + \vec{\omega}_2 =\vec{\omega}_3 + \vec{\omega}_4 \neq 0$,
the coupling functions
do not flow in the leading order in $\frac{1}{\lambda}$.
But for sector indices
$\vec{\omega}_1 + \vec{\omega}_2 =\vec{\omega}_3 + \vec{\omega}_4 = 0$,
the coupling functions, $g^{(0)}_{BCS}(\vec{\omega}_1,\vec{\omega}_4)$,
flow.
The diagrams that contribute to the flow equation
are the {\bf ladder diagrams} with
self energy insertion for the internal electron lines but with no other
two legged subdiagram.The amplitude of such a diagram with
$n$ interaction squiggles and with zero incoming and outgoing
box momenta of the particles is given by,
\begin{eqnarray}
\big( \frac{1}{\lambda^{d-1}} \big)^{n+1}
\sum_{\vec{\omega_1},...,\vec{\omega_n}}
(-1)^n \beta^n g_{BCS}(\vec{\omega},\vec{\omega}_n)
g_{BCS}(\vec{\omega}_n,\vec{\omega}_{n-1}) ...
g_{BCS}(\vec{\omega}_1,\vec{\omega}^{\prime})  \nonumber
\end{eqnarray}
In the equation above, $(\vec{\omega}^{\prime},-\vec{\omega}^{\prime})$
and $(\vec{\omega} ,-\vec{\omega})$ are sector indices of incoming
and outgoing electron lines respectively.Other sector indices correspond
to the internal electron lines. $\beta$ is a strictly positive number
coming from the fermion loop integration in the  Feynman diagram and is
given by
$ ~\beta =\int dk_{\perp}dk_{\parallel}dk_{0}
 ~[k_{0}^{2} +(v_F k_{\parallel}-\lambda \delta\mu_{1})^{2}]^{-1} ~$.
We find that the renormalized value of $g_{BCS}$ is
\begin{eqnarray}
g_{BCS}^{(j+1)}(\vec{\omega} , \vec{\omega}^{\prime})
&=&g_{BCS}^{(j)}(\vec{\omega} , \vec{\omega}^{\prime})
+\sum_{n=1}^{\infty}\big( \frac{1}{\lambda^{d-1}} \big)^{n}
\sum_{\vec{\omega_1},...,\vec{\omega_n}}
(-1)^n \beta_{j}^{n} g_{BCS}^{(j)}(\vec{\omega},\vec{\omega}_n)
g_{BCS}^{(j)}(\vec{\omega}_n,\vec{\omega}_{n-1}) ...
g_{BCS}^{(j)}(\vec{\omega}_1,\vec{\omega}^{\prime}) \nonumber\\
& & +O(\frac{g^{(j)}}{\lambda_j})
\end{eqnarray}
The explicit expression for flow equation can be obtained by
expanding  the coupling functions
$g_{BCS}(\vec{\omega},\vec{\omega}^{\prime})
=g_{BCS}(\langle \vec{\omega},\vec{\omega}^{\prime})$
into spherical harmonics,
$g_{BCS}(\langle\vec{\omega},\vec{\omega}^{\prime})
=\sum g_{l}h_{l}(\langle\vec{\omega},\vec{\omega}^{\prime})$.
Up to terms of order $\frac{1}{\lambda}$,
\begin{eqnarray}
g_{l}^{(j+1)}= \frac{g_{l}^{(j)}}{1+\beta_{j}g_{l}^{(j)}}
+O\big(\frac{1}{\lambda}\big)
\end{eqnarray}
To obtain the differential equation for the R.G. flow, let us define
$g_{l}^{(j)}$ := $g_{l}^{(\lambda_j)}$, and consider a scale
$\lambda=e^{t}\lambda_0$. Next,
define $g_{l}(t)$:=$g_{l}^{(e^t \lambda_{0})}$. The coefficient
$\beta=\beta(t,t')$ vanishes in the limit $t' \searrow t$, therefore,
$\beta(t',t)=(t'-t)\gamma(t) +O\big( (t'-t)^{2}\big)$. Writing difference
equation for the couplings $g_{l}^{(j+1)}$ and $g_{l}^{(j)}$, and
dividing both sides of the difference equation by $(t'-t)$,
we finally obtain
\begin{eqnarray}
\frac{d}{dt}g_{l}(t)=-\gamma g_{l}(t)(t)^{2} +
O\big(e^{-t}g(t)^{2}\big)
\end{eqnarray}
where $ ~\gamma=\gamma(t)>0 ~$.
It is independent of $l$ and approximately
independent of $t ~$, and therefore, we set $\gamma=\gamma_{0}$.
The positivity of $\gamma$ follows from slow monotone growth of
$\beta (t'-t)$ in $t'$.
Neglecting the error term, the solution can be written as,
\begin{eqnarray}
g_{l}(t)=\frac{g_{l}(0)}{1+\gamma_{0}g_{l}(0)t}.
\end{eqnarray}
If the coupling constants are positive or rather non-negative,
$g_{l}(0)\geq 0$, the effective
running coupling constant goes to zero,
and we have the Landau-Fermi liquid
phase. This phase consists of free quasi-particles which are
electrons and holes with renormalized mass and the chemical potential.
On the other hand,
if there is an angular momentum channel,$ ~l ~$, with attractive
interactions ($g_{l}(0)<0$) the flow diverges at a finite value
of the scaling parameter, $t=-(\gamma_{0}g_{l}(0))^{-1}$ .
This singularity reflects the instability
of the ground state.\\
The ground state of the perturbation theory described above was
taken to be the non-interacting Fermi gas with Fock space constructed
from the elementary excitations, electrons and holes.
The renormalization group analysis shows that this
is not the true ground state in the presence of atrractive
interaction. The singularity in the running coupling constant
reflects just this fact. To see
whether the true ground state is a BCS state, we need to know the nature
of the pole. It is easily found that the
residue at the pole of the effective running coupling constant
has negative sign. This
signifies the presence of Cooper pairs \cite{shirkov,abrik}. The true
ground state is thus, the BCS ground state of superconductivity.
In the following section, we shall see that the Landau
pole, which looks very similar to the pole of the effective coupling
constant in condensed matter system, has positive residue at the pole.
This sets it apart from the BCS type of instability \cite{shirkov,abrik}.

\section{Dyson's arguments for the divergence of
perturbation series in coupling constant}

Dyson's arguments for the divergence of perturbation theory in QED
is elegant in its' simplicity.
We will simply reproduce here his arguments \cite{dys}.\\
Let us suppose that
\begin{eqnarray}
F(e^2)=a_0+a_1 e^2 +a_2 e^4+...              \nonumber
\end{eqnarray}
is a physical quantity which is calculated as a formal power series in
$e^2$ by integrating the equations of motion of the theory
over a finite or infinite time. Suppose that the series
converges for some positive small value of $e^2$; this implies that $F(e^2)$
is an analytic function of $e$ at $e=0$. Then for sufficiently small value
of $e$, $F(-e^2)$ will also be a well-behaved analytic function with
a convergent power series expansion.
However, for $F(-e^2)$ we can also make a physical interpretation.
In the fictitious world, like charges attract each other.The potential
between static charges, in the classical limit of large distances
and large number of elementary charges, will be just the Coulomb potential
with the sign reversed. But it is clear that in the fictitious world
the vacuum state as ordinarily defined is not the state of lowest energy.
By creating a large number $N$ of electron-positron pairs, bringing the
electrons in one region of space and the positrons in
another separate region, it is easy to construct a pathological state
in which the negative potential energy of the Coulomb forces is much
greater than the total rest energy and the kinetic energy of the particles.
Suppose that in the fictitious world the state of the
system is known at a certain time to be an ordinary physical state with
only a few particles present.
There is a high  potential barrier separating
the physical state from the pathological state of equal energy.
However, because of the quantum mechanical
tunneling effect, there will always be a finite probability that in any
finite time-interval the system will find itself in a pathological state.
Thus every physical state is unstable against the spontaneous
creation of many particles. Further, a system once in a pathological state
will not remain steady; there will be rapid creation of more and more
particles, an explosive disintegration of the vacuum by spontaneous
polarization. In these circumstances it is impossible that the
integratation of the equation of motion of the theory over any finite
or infinite time interval, starting from
a given state of the fictitious world, should lead to well-defined
analytic functions.Therefore $F(-e^2)$ can not be analytic and the series
can not be convergent.\\
 The central idea in Dyson's arguments for the divergence
of perturbation theory in coupling constant, as is evidient
from the lengthy discussion above, is that the convergence of the
perturbation theory in coupling constant would lead to the
existence of pathological states to which the normal states of
QED would decay.These pathological states correspond to
states of a  quantum field theory whose vacuum state is unstable.
Therefore, if quantum electrodynamics is  a meaningful theory,
the perturbation series must diverge
(for more discussions related to these arguments,
see \cite{ark,dunne} and references there in ).\\
It should be noted that Dyson's proof appeared
much before the advent of asymptotic freedom in quantum field
theories. The main point in Dyson's proof is that,
if the perturbative series is convergent, then for small
value of $e^2$ , we can analytically continue to $-e^2$ and then this
series will also be convergent. Let us consider the series for the
vacuum polarization (two point Green's function for photons). Both the
perturbative as well as the analytically continued series
are assumed to be converegent. We can carry out loop wise summation.
We can write the formal sum for one loop
diagrams ( the coupling is assumed to be small)
and extract from it the effective coupling constant. In the analytically
continued theory, we obtain
\begin{eqnarray}
e^{2}_{eff}= \frac{e^2}{1-(-e^2)ln\frac{\Lambda^2}{k^2}}
\end{eqnarray}
When $\Lambda \rightarrow \infty$, then ~~$e^{2}_{eff}\rightarrow 0$~~, and
therefore the analytically continued theory is asymptotically free.
It is also easy to infer that at low energies the effective coupling
constant increases.
On purely formal grounds, the efective coupling constant of this purely formal
theory behaves in the same way as the effective coupling constant
in QCD for both the high energy and low energy limits inspite of the
fact that in the formal theory interactions are mediated by abelian
gauge fields. For further considerations,
we require an infrared regulator. We can
always choose an infrared regulator such that the effective
coupling constant remains small. It is not hard to see now that the
classical argument of Dyson holds. Let us consider a box of size $L$
with wave functions vanishing at the boundary. The size, $L$, itself
can be taken as the infrared regulator. One may suspect that, as the
the number of pairs of electron-positron created, as per Dyson's
arguments, increases, the inter-particle distance decreases and this
leads to decrease of $e_{eff}^2$, and since $e_{eff}^2$ goes to zero
as the inter-particle distance goes to zero ( because of asymptotic
freedom) , the vacuum state somehow stabilizes. However, it turns
out that the suspision is unfounded, and the reason behind this is
that the effective coupling constant, $e_{eff}^2$,
decreases too slowly with the decrease in the inter-particle distance.
Let us suppose that $N$-pairs of electron-positrons are created, the
electrons separate to one half of the box and the positrons to the
other half. In this process, the vacuum energy decreases
at least by $E_0\sim -N^2 e_{eff}^2$
( other factors are suppressed ).
Now $e_{eff}^2$, for large $N$, goes as, $e_{eff}^2\sim 1/ln(N^{1/3})$.
Therefore, the decrease in the vacuum energy by the process of pair
creation goes as $E_0\sim -N^2 e_{eff}^2=-N^{2}/ln(N^{1/3})$, which
approaches $-\infty$ as $N\rightarrow \infty$. This demonstrates that
the vacuum energy remains unbounded below. Therefore,
asymptotic freedom does not change the pathological character of
the analytically continued theory and Dyson's argument
remains intact.\\

\section{Landau singularity and the vacuum state}

The Lagrangian of QED with number of flavours, $N_{f}$, is
given by \cite{azam},
\begin{equation}
{\cal L} =  \sum_{j=1}^{N_f} \bar{{\psi}}^j \Big(i\gamma^{\mu} \partial_{\mu}
+ m - e \gamma^{\mu} A_{\mu}\Big) \psi^{j} + \frac{1}{4} F_{{\mu}{\nu}}^2
\end{equation}
where $ \psi^{j}$  and $\bar{{\psi}}^j$
are the   Dirac field and its' conjugate, $j$ is the flavour index,
and $ A_{\mu}$  and
$F_{{\mu}{\nu}}$ are the electromagnetic
potential  and the field strength respectively. We will
investigate this model when the number flavours,
$N_f$, has both the positive and negative sign. We,
therefore, introduce the notation $|N_f|=sign(N_f)\times N_f$~.
For the Lagrangian, written above,  ~$\frac{1}{N_f}~$ expansion
is introduced by assuming that, in the limit
~$|N_{f}|{\rightarrow}{\infty}$, ~$e^{2}|N_{f}|~
=constant= ~\alpha^{2} ~(say)~$ .
Large flavour limit of quantum electrodynamics can be equivalently
described by the Lagrangian,
\begin{equation}
{\cal L} =  \sum_{j=1}^{N_f} \bar{{\psi}}^j \Big(i\gamma^{\mu} \partial_{\mu}
+ m - \frac{e}{\sqrt |N_f|}~ \gamma^{\mu} A_{\mu}\Big) \psi^{i} +
\frac{1}{4} F_{{\mu}{\nu}}^2
\end{equation}
With this form of the Lagrangian, it is easier to set up Feynman diagram
technique.To each photon and fermion line corresponds their usual propagator.
Each vertex contributes a factor of $\frac{e}{\sqrt{|N_f|}}~$, each fermion
loop contributes a factor of $~(-1)~$ for anticommuting
fermions and a factor of
$~N_f~$ because of summation over fermion flavours.Using these rules, it is
easy to set up $~1/N_f~$ expansion series for any physical observable.
Just as in the case of perturbation theory in the coupling constant, the
expansion  in $~\frac{1}{N_f}~$ allows us to express an observable $~F~$
in the form,
\begin{eqnarray}
F(\frac{1}{N_f})=Q_{0}+\frac{1}{N_f}~Q_{1}+\frac{1}{N_{f}^{2}}~Q_{2}+... ...
\end{eqnarray}
$Q_{0}~,~Q_{1}~,~Q_{2},~... ...~~$ are some  functions of the coupling contant.
Now suppose
that the series converges for some small value of $\frac{1}{N_f}~$ ( large
value of $N_f~$ ), then the observable function
$F(\frac{1}{N_f})$ is analytic for
$\frac{1}{N_f}=0~$ ( $N_f=\infty~$ ).Therefore, we can consider a small
negative value of $~\frac{1}{N_f}~$ ( large negative value of $N_f~$ ) for
which the function is analytic and convergent.In other words, the function
$~F(\frac{1}{N_f})~$ can be analytically continued to small negative value
of $~1/N_f~$ and the series thus obtained will be convergent.
Latter, in  the text, we will discuss the meaning of negative flavour.
Here we just mention that, in the context of lattice QCD,
fermions with finite number of negative flavours
have been considered before(see \cite{divi,rolf,anthony} and
references there in).\\
Let us calculate the effective coupling constant
from the formal $~1/N_f~$ expansion series of the
two point Green's function.
The series is assumed to be convergent, and therefore, for sufficiently
large $~N_f~$, one can restrict to the leading order term.The leading
order term is given by the one-loop diagrams which can
easily be eavaluated to obtain the polarization from which one can
read off the effective coupling constant. It is given by,
\begin{equation}
e_{eff}^{2}(\Lambda^2) = {e^2  \over 1- \frac{e^2 N_f}{3\pi |N_f|}
ln \frac{\Lambda^2}{m^2}}
\end{equation}
If $N_f$ is positive,
\begin{equation}
e_{eff}^{2}(\Lambda^2) = {e^2  \over 1- \frac{e^2}{3\pi}
ln \frac{\Lambda^2}{m^2}}
\end{equation}
It is to see that the effective coupling constant has a pole at finite
but very large value of $\Lambda^2=m^2~exp(3\pi^2/e^2)~$. This is
known as the Landau singularity of the effective coupling constant in
QED. This is the central theme of the paper, and therefore, we repeat
that the convergence of the $1/N_f$ expansion series allows us to choose
a sufficiently large $N_f$
and restrict to the leading order terms in $1/N_f$.\\
Thus :-
\begin{itemize}
\item{The appearance of Landau singularity in the effective running
coupling constant in QED is intimately linked
to the assumption that the $1/N_f$ expansion series converges.}
\end{itemize}
We will now argue that Landau singularity leads to the instability
of vacuum state in QED. We mentioned before that, the
convergence of the series for
sufficiently large positive $N_f$, allows us to analytically
continue it to negative $N_f$, and the series will again be convergent.
Thus, for $N_f$ negative, we obtain,
\begin{equation}
e_{eff}^{2}(\Lambda^2) = {e^2  \over 1+ \frac{e^2}{3\pi}
ln \frac{\Lambda^2}{m^2}}
\end{equation}
From this equation, it is easy to see that in the limit
$\Lambda\rightarrow\infty~$, $e_{eff}^{2}\rightarrow 0$.
Therefore, the formal theory that we obtain from the
analytical continuation of $1/N_f~$ ( ~for large $N_f$~)
to the small negative
value of $~1/N_f~$, is ( at least formally ) asymptotically free.This
seem to suggest that the physical meaning of the negative sign of
$N_f$ could possibly be traced in the free theory without the interaction
term. In following sections, we shall argue that the  choice
of negative $N_f$ for anticommuting fermions amounts to considering
commuting fermions with positive $N_f$.\\
For the Lagrangian
given by Eq.(25) (in four dimensional Euclidean space),
the partition function is given by functional integral,
\begin{eqnarray}
{\cal Z}_{ac}=\int DA(x) D\bar{{\psi}}(x) D\psi(x) exp(-\int d^{4}x{\cal L})
\end{eqnarray}
Funtional integration with respect to the anticommuting
fermion fields (grassman variables) gives,
\begin{eqnarray}
{\cal Z}_{ac}=&&\int DA(x)~ det^{N_f}(i\gamma^{\mu} \partial_{\mu}
+ m - e \gamma^{\mu} A_{\mu}) \nonumber\\
&&exp(- \frac{1}{4} \int d^{4}x  F_{{\mu}{\nu}}^2)
\end{eqnarray}
On the other hand, if we consider the fermion fields to be commuting
variables, then the integration above ammounts to functional
integration over complex fields, and we obtain,
\begin{eqnarray}
{\cal Z}_{c}=&&\int DA(x)~ det^{-N_f}(i\gamma^{\mu} \partial_{\mu}
+ m - e \gamma^{\mu} A_{\mu}) \nonumber\\
&&exp(- \frac{1}{4} \int d^{4}x  F_{{\mu}{\nu}}^2)
\end{eqnarray}
Note that this expression could be obtained from the previous expression,
simply by assuming that  $N_f$ is  negative.Therefore, anticommuting
fermions  with negative $N_f$ has the same partition function as
the commuting fermions with positive $N_f$. Since, physically
interesting observables can be calculated from the partition function,
our claim is that the negative flavour anticommuting fermions are
equivalent to the positive flavour commuting fermions.
\par We can also arrive at this conclsion using
the Feyman diagram tecnique of the formal perturbation theory.
Consider the two point Green's functions for the photons using Lagrangian
given by Eq.(26) . First, we consider just one loop diagram and show how the
contribution due to flavours appears in the calculations.
There are two vertices and a fermion loop,
each vertex contributes a factor of $~\frac{e}{\sqrt{|N_f|}}~$,
the fermion loop contributes a multiplicative factor of $~(-1)~$
because the fermions anticommute
and a multiplicative factor of  $N_f$ because of summation
over flavours of the internal fermion lines.Now, if $N_f$ happens to be
negative, the factor $~(-1)~$ and  the factor $N_f~$, combines to give the
factor $~|N_f|~$.This is also the contribution if  the fermions commute and
the flavour is positive ( the factor $~ (-1)~$ is
absent for commuting fermions).
The same procedure applies for the multiloop  diagrams.
This shows that the choice
of negative $~N_f~$ for anticommuting fermions amounts to considering
commuting fermions with positive $~N_f~$.
We have shown earlier in text
that quantum electrodynamics with anticommuting fermions and negative value
of $~N_f~$ is asymptotically free. Therefore, our purely formal
quantum electrodynamics with commuting fermions and positive $~N_f~$ is
asymptotically free \cite{azam}.
It is well known that the quantum field theory of free commuting
fermions does not have stable vacuum state \cite{gel,nai,wein}.
It then follows that
an interacting asymptotically free quantum field theory
built around such a vacuum
state can not be stable: all states in this theory will be pathological.
These results follow from the single assumption
that the $1/N_f$-expansion series in QED is convergent, and therefore
analytic in $1/N_f$. As per Dyson's argument, this would lead
to the decay of normal states of QED with anticommuting
fermions to the pathological states of QED with
commuting fermions via the process of quantum mechanical tunnelling.
Therefore, the vacuum state of QED with large number of flavour of
anticommuting fermions can not be stable.\\
We have already seen that the convergence of $1/N_f$ expansion theory
invariably leads to the appearance of Landau singularity.\\
Thus :-
\begin{itemize}
\item{Landau singularity signals the instability of vacuum state
of quantum electrodynamics.}
\end{itemize}
We have, no where, in text shown that the $1/N_f$ expansion
series diverges. It was only an assumption.
There exist large number of publications which, based on the
behaviour of the large order terms in the series (in coupling constant),
suggest that the perturbation series is, possibly, divergent.
Similar arguments can be extended to our case unless there is
some magical cancellations in large orders of the series.
But such magical cancellations, if any, would plague the
theory with the instability of vacuum state and render it
meaningless.\\
Thus :-
\begin{itemize}
\item{Quantum electrodynamics with large number of flavours of
(anticommuting) fermions will be a meaningful theory, only if
the  series in $~1/N_f~$ expansion diverges.}
\end{itemize}
In the abstract, we annouced the extraordinary success of
quantum electrodynamics. How do we understand this success
in the light of results obtained in this paper?
In the following section, we discuss how a divergent series
could possibly lead to a meaningful theory of quantum electrodynamics.

\section{A brief mathematical digression:asymptotic expansion series}

There is an imporatant class of series, known as asymptotic series,
which frequently appear in physical problems.
These series behave like a convergent
series upto a certain number of terms but after that it behaves like
a divergent series. This type of series is called asymptotic series
\cite{siro,bend} and is generally defined through a
power series represetation of a function.
The behaviour of an asymptotic series is very transparent in
the following example (a version of Stirling's formula) given
by Bender and Orszag (page 218 in \cite{bend})
for the asymptotic series expansion of factorial,
$N!=~\big{(}Z-1\big{)}!~$, in  powers of $~1/Z~$.

\begin{eqnarray}
&&(Z-1)!=(2\pi/Z)^{1/2}e^{-Z}Z^{Z}\Bigg{(}~ 1+\frac{1}{12}Z^{-1}+
\frac{1}{288}Z^{-2}-\frac{139}{51840}Z^{-3} -\frac{571}{2488320}Z^{-4}+
\nonumber\\
&&\frac{163879}{209018880}Z^{-5}+\frac{5246819}{75246796800}Z^{-6}-
\frac{534703531}{902961561600}Z^{-7}
-\frac{4483131259}{86684309913600}Z^{-8}+...\Bigg{)}       \nonumber
\end{eqnarray}
For example, for $~0!~$, the terms get smaller for a while but the
$15th$ term becomes larger than $0.01$, and the $35th$ is bigger than
$10^{10}$. On the other hand, the $35th$ term for $9!$ is
$10^{10}/10^{35}$ which is quite small but the $175th$ term is
bigger than one, and the $199th$ term is nearly $10^{12}$.
This series has zero radius of convergence. A non-zero
radius of convergence in $1/Z$ would also include some
negative values around zero, and therefore, if the series
for $~\big{(}Z-1\big{)}!~$ converges for some large positive integer
$~Z=\big{(}N+1\big{)}~$, it will also converge for some large negative integer
$~Z=-\big{(}N+1\big{)}~$.
But $\big{(}Z-1\big{)}!=~\big{(}-N-2\big{)}!~$  is infinity.
Therefore, the
the $~1/Z~$ expansion series of $~\big{(}Z-1\big{)}!~$
for $~Z=N+1~$ can not converge. The series
is meaningful only as an asymptotic expansion.\\

Mathematically, a function $f(x)$ is said to have an asymptotic power series
represetation if for all $n$,
\begin{eqnarray}
\lim_{x\rightarrow 0}|\frac{f(x)-\sum_{i=0}^{n} a_i x^i}{x^n}|=0
\nonumber
\end{eqnarray}
In other words,
\begin{eqnarray}
f(x)=\sum_{i=0}^{n} a_i x^i +{\mathcal{O}}(x^n) \nonumber
\end{eqnarray}
This means that the error in estimating the function is of the same
order as the last term in the series.
To explain, let us consider the following function,
\begin{eqnarray}
F(x)=\int_{0}^{\infty}\frac{e^{-t}}{1+xt}dt
\nonumber
\end{eqnarray}
for real positive $x$ and $x\rightarrow 0$.
Since,
\begin{eqnarray}
\frac{1}{1+xt}=1-xt+x^2 t^2+...+\frac{(-xt)^k}{1+xt}
\nonumber
\end{eqnarray}
we have,
\begin{eqnarray}
F(x)=\sum_{k=0}^{N} (-1)^k x^k k! +R_{N+1}(x)~~;~~|R_N(x)|=N! x^N
\nonumber
\end{eqnarray}
The ratio of the  two successive terms is
\begin{eqnarray}
\frac{x^k k!}{x^{(k-1)} (k-1)!}=xk
\nonumber
\end{eqnarray}
This shows that the terms first decrease (since by assumption
$0<x<<1$) and then increase (when $k>\frac{1}{x})$.
From this it follows that for a given value of $x$, there exists
a best approximation. In other words, for a fixed value of
$x$, only a definite accuracy can be achieved. However, the function
defined by integral is well behaved and is non-analytic in $x$ at $x=0$.\\
We have seen in previous sections that perturbation series of quantum
eletrodynamics in coupling constant as well as in $~1/N_f~$ is divergent.
Studies of the large order terms in perturbation series suggest
that these series are probably asymptotic in nature. The fine structure
constant of QED is $1/137$
which is quite small, and therefore in the asymptotic series, one can
consider terms up to a very large order. This can possibly explain
the spectacular success of QED, in spite of fact that the series
is divergent and asymptotic.\\
In our view it does not make sense to look for any kind of singularity
in coupling constant of a theory with asymptotic expansion and try
to draw any big physical conclusion.
There are attempts to give meaning to this kind of series through Borel
summation but, again in our view, one does not achieve more than
what one obtains through the summation of asymptotic series as
explained above.
\section{ Conclusions}
In this paper, we have discussed several aspects of perturbation
theory in quantum electrodynamics. Most importantly,
we have shown that the Landau singularity appears in the leading
order terms in the $1/N_f$ expansion series. The restriction
to leading order terms makes sense only when the series converges.
We used Dysonian argument to show that the convergence
of the series leads to the instability of vacuum state in QED.
This demonstrates that Landau singularity reflects the instability
of vacuum state. These problems can be avoided if the series
diverges. Divergent series as asymptotic series can provide
physically meaningful results within some unavoidable errors
depending on the value of the expansion parameter.
The fine structure constant, which is the expansion parameter in
QED, is small, and therefore, there is no surprise why QED
is such a successful theory.\\
Divergence of the perturbation series, suggested by Dysonian
argument, is to save the vacuum state from the catastrophic
disintegration. But that is not the sole point of Dysonian argument.
It also suggests non-analyticity of observables as a function
of the coupling constant, $e^2$ when $e^2=0$ ( similarly in $1/N_f$
for $N_f=\infty$). Note that the non-analyticity in
coupling constant seems to be in the infrared (IR) region.
On the other hand, the non-analyticity of $1/N_f$ expansion
series for $N_f=\infty$ is connected with Landau
singularity which is in the ultrviolet (UV) region.
It is not clear how the IR and UV regions are connected.
However, we must remember that we are dealing with pathological
situations where there are singularities and divergences.
The main point of Dysonian argument, in our view,
is that the non-perturbative QED
is non-analytic and this behaviour can not be captured by a
perturbative power series. More than fifty years of theoretical
research has not been able to find a non-perturbative formulation
which can remove the pathological aspects of QED. May be it is
time for experimentalists to step in and look for ways to
find non-analytical dependence of some physical observales
on the coupling constant or flavours.
\section{Acknowledgements}
I would like to thank here a large number of friends and colleagues
with whom I discussed the problem of Landau singularity for almost
a quarter century, and I can not possibly recall names of all of them.
Special thanks are due to Prof. Spenta Wadia from whom I learnt about
the Landau singularity and Dyson's proof of divegence of
perturbation theory in the year 1983 when I was a graduate student
in Tata Institute of Fundamental Research, Mumbai, India.
I would also like to use this opportunity to thank
Prof. N.V. Mitskievich, my M.Sc. thesis adviser and Prof. P.P.
Divakaran, my Ph.D. thesis advisor, from whom I learnt various aspects
of classical and quantum field theories.


\end{document}